\documentclass{sigchi}

\usepackage{balance}  
\usepackage{graphics} 
\usepackage{times}    
\usepackage{url}      
\usepackage{algorithm}
\usepackage{algorithmic}

\makeatletter
\def\url@leostyle{%
  \@ifundefined{selectfont}{\def\UrlFont{\sf}}{\def\UrlFont{\small\bf\ttfamily}}}
\makeatother
\def\pprw{8.5in}
\def\pprh{11in}

\usepackage[pdftex]{hyperref}
\hypersetup{
pdftitle={SIGCHI Conference Proceedings Format},
pdfauthor={LaTeX},
pdfkeywords={SIGCHI, proceedings, archival format},
bookmarksnumbered,
pdfstartview={FitH},
colorlinks,
citecolor=black,
filecolor=black,
linkcolor=black,
urlcolor=black,
breaklinks=true,
}

\newcommand\tabhead[1]{\small\textbf{#1}}

\begin{document}
\title{A Ranking Algorithm for Re-finding}

\numberofauthors{1}
\author{
  \alignauthor Gangli Liu\\
    \affaddr{Department of Computer Science}\\
    \affaddr{Tsinghua University}\\
    \email{gl-liu13@mails.tsinghua.edu.cn}\\
}
\maketitle

\begin{abstract}
Re-finding files from a personal computer is a frequent demand to users. When encountered a difficult re-finding task, people may not recall the attributes used by conventional re-finding methods, such as a file's path, file name, keywords etc., the re-finding would fail.

We proposed a method to support difficult re-finding tasks. By asking the user a list of questions about the target, such as a document's pages, author numbers, accumulated reading time, last reading location etc. Then use the user's answers to filter out the target.

After the user answered a list of questions about the target file, we evaluate the user's familiar degree about the target file based on the answers. We devise a ranking algorithm which sorts the candidates by comparing the user's familiarity degree about the target and the candidates.

We also propose a method to generate re-finding tasks artificially based on the user's own document corpus.
\end{abstract}



\section{Introduction}

As computers are becoming an indispensable tool for people, re-finding a file which has been accessed previously becomes a common task \cite{Jensen2010}.  Some of the re-finding tasks are effortless to accomplish, especially when the user is familiar with the target file or the file has been accessed recently. Other re-finding tasks are difficult due to the user's vague memory about the target.

We proposed a method to support difficult re-finding tasks in our previous works\cite{liu2016method}.
Based on the cognitive psychology finding, we propose to use a question and answer wizard interface to collect all the possible memories of the user about the target document.

To alleviate the user's cognitive burden, we try to give recommendations for each question of the list, the recommendations are generated based on analyzing the database which contains each file's attributes and the user's experience about each file.

Besides acting as a memory collector and a filter to exclude irrelevant files, the question list also can be used as an estimator to evaluate how familiar the user is about the target file. If the user answered the questions quickly or even gave more precise answers than the recommendations, we can infer that the user is familiar with the target file; In contrast, if the user answered slowly and skipped many questions because s/he did not know an answer, we can infer that he is unacquainted about the target file. Based on these factors we can compute a familiarity degree about the target file.

One question is how we can estimate the user's familiarity degree about each candidate file. As we have a database which records the user's experience about each file, such as access frequency, accumulated process time, last access date etc. We can use these data to computer familiarity degrees about each candidate file. By comparing these familiarity degrees, we can produce a ranking algorithm to rank the still remaining candidates.

Figure 1 shows the framework of our method.

\begin{figure}[!h]
\centering
\includegraphics[width=1.0\columnwidth]{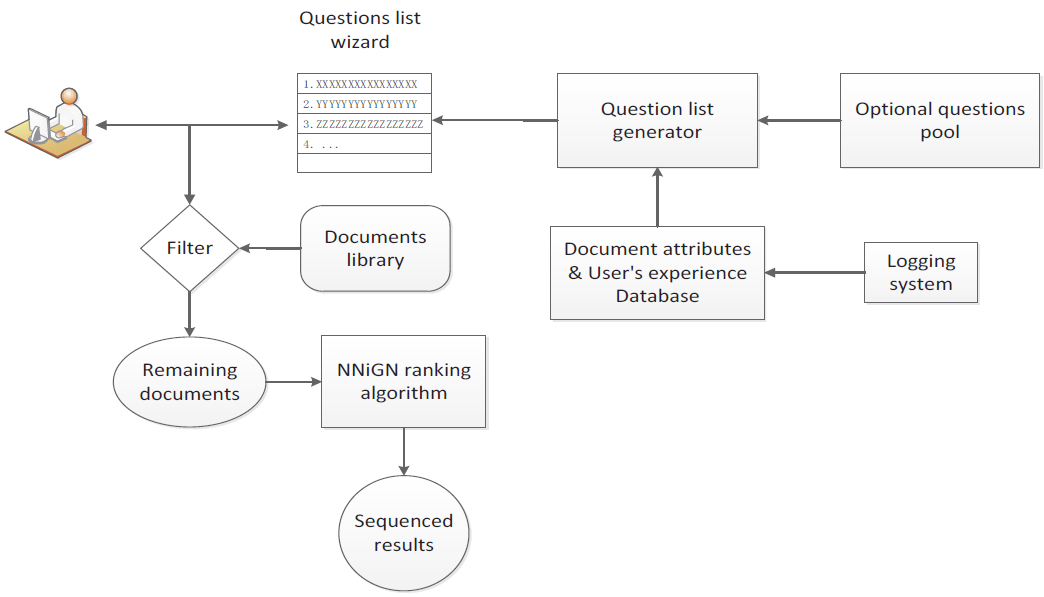}
\caption{The framework of the method.}
\label{fig:figure1}
\end{figure}

\section{A Question and Answer Wizard}
A question and answer wizard can help the user recall his/her memories about the target document. Generally, these memories can be classified into two categories, one category includes those attributes possessed by a document inherently, such as its author, size, path, file name, keywords etc. The other category includes a user's experiences about a document, such as access times, print experience, discussion with someone about the document, a document's provenance etc. We classify inherent attributes into two categories too: metadata and extended metadata. Figure 2 shows a classification of the items which are used by our method.
\begin{figure}[!h]
\centering
\includegraphics[width=1.0\columnwidth]{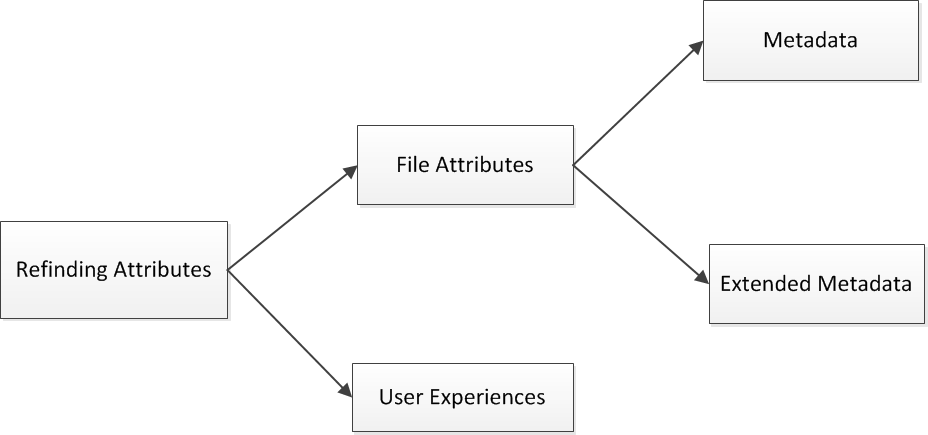}
\caption{Classification of re-finding attributes.}
\label{fig:figure2}
\end{figure}

\subsection{Metadata}

File metadata includes those attributes which can be easily extracted and are provided by the operating system, we do not need to develop new methods and tools to obtain them. Such as File size, File type, Last modified date, Last access date, Date of creation etc.

\subsection{Extended Metadata}

Extended metadata are those attributes which are not easy to obtain. Although they inherently belong to a document, but are not provided by the operating system. Some of them can be extracted by existing tools or methods, like a document's keywords, topic, and pages. For others, there are no existing methods or tools to extract them, such as how many pictures a document contains, the genders of a document's authors. To obtain them, we need to develop new methods and tools. Table 1 shows some of them.

\begin{table}
  \centering
  \begin{tabular}{|c|c|}
    \hline
    \tabhead{Name} &
    \multicolumn{1}{|p{0.6\columnwidth}|}{\centering\tabhead{Description}}\\
    \hline
    Number of authors &\\
    \hline
    Genders of authors &\\
    \hline
    Pages & Page size of the document\\
    \hline
    Number of images &  \\
    \hline
    Number of tables &  \\
    \hline
    Color of images &  Monochrome or colorized\\
    \hline
    Content category & Novel, news or research paper etc. \\
    \hline
    Difficulty level & How difficult to understand it? \\
    \hline
    Topic & What is it about? \\
    \hline
    Language & In what language it is written? \\
    \hline
    Bibliography & Does it contain a bibliography?\\
    \hline

  \end{tabular}
  \caption{Attributes of Extend Metadata.}
  \label{tab:table1}
\end{table}

\subsection{User Experiences}

Another category of information we can collect and use is the user's experiences about each document. As extended metadata, some of them are easy to obtain, like how many times the user has accessed a document; some of them are not so easy to obtain, such as whether the user is familiar with its authors, association with the user's own works. Table 2 shows some of them.
\begin{table}
  \centering
  \begin{tabular}{|c|c|}
    \hline
    \tabhead{Name} &
    \multicolumn{1}{|p{0.5\columnwidth}|}{\centering\tabhead{Description}}\\
    \hline
    Provenance & Where is it from?\\
    \hline
    Printing experience & Whether printed it before\\
    \hline
    Re-finding experience & Whether retrieved it before\\
    \hline
    User association & Association with \\
                     &the user's own works  \\
    \hline
    Author familiarity & How familiar about its authors? \\
    \hline
    Unusual access time & Whether a night or weekend?\\
    \hline
    Unusual access location & Whether at an unusual place? \\
    \hline
    Access frequency & \\
    \hline
    Accumulative process time & \\
    \hline
    Coverage & Have read fully or only a part? \\
    \hline

  \end{tabular}
  \caption{Attributes of user experiences.}
  \label{tab:table2}
\end{table}

\subsection{The Logging and Analyzing System}

To collect all these information that might be useful is a huge project. Some of the attributes can be collected by analyzing the document separately, some of them need to be collected during the user is processing a document, such as \textit{``Accumulative process time"} or \textit{``Coverage of the document"}. For many of them, we have not existing tools to extract them, we need to devise new methods and tools to obtain them. Figure 3 shows a paradigm of the logging and analyzing system.
\begin{figure}[!h]
\centering
\includegraphics[width=1.1\columnwidth]{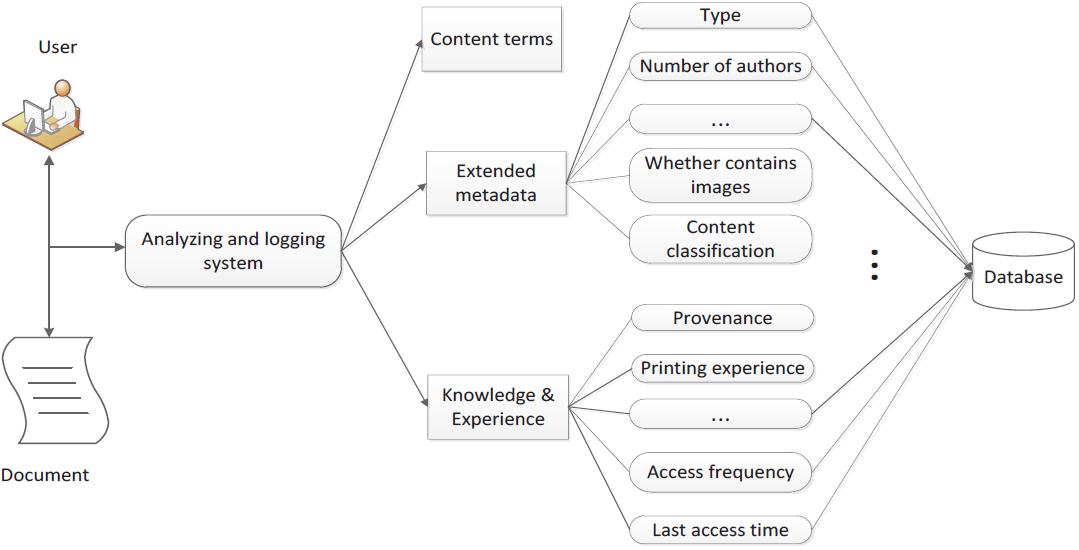}
\caption{A paradigm of the logging and analyzing system.}
\label{fig:figure3}
\end{figure}

\section{A Rank-by-familiarity Ranking Algorithm}

After the user has answered all the questions and we used these answers excluded all unrelated documents. There may still be some candidates remained, the quantity is large enough that the user cannot use a scan and recognition strategy to locate the target. Now we need a ranking method to sort the remaining candidates. As mentioned above, we can compute the user's familiarity degree about the target document and each candidate document, then compute their differences, the smaller the difference is, the higher the candidate will be ranked.

This ranking algorithm can work even if the user knows little about the target document, because under this condition those documents which the user is unacquainted will be ranked on a higher position. Therefore, we call it the No-News-is-Good-News(NNiGN) ranking algorithm.
\subsection{How to estimate a user's familiarity degree about a document?}
Just by intuition, the more time a user used to read a document, or the interval between his/her last access of the document is shorter, or the document's topic is closer to the user's profession or interest, the more familiar the user will be about the document. So we propose to use these quantities and linear regression model to estimate a user's familiarity degree about a document.

Table 3 shows the variables we use to calculate the familiarity degree and their symbols.
\begin{table}
  \centering
  \begin{tabular}{|c|c|}
    \hline
    \tabhead{Variable} &
    \multicolumn{1}{|p{0.2\columnwidth}|}{\centering\tabhead{Symbol}}\\
    \hline
    Familiarity degree to candidate i & $F_i$\\
    \hline
    Frequency of visiting candidate i & $R_i$\\
    \hline
    Cumulative time of processing candidate i & $C_i$\\
    \hline
    Interval from last access & $I_i$  \\
    \hline
    Distance from the user's profession or interest & $D_i$ \\
    \hline
  \end{tabular}
  \caption{Variables used to estimate document familiarity degree.}
  \label{tab:table3}
\end{table}
We use the following linear regression formula to calculate the familiarity degree about the i-th document:
\begin{equation}F_i = \alpha_0 + \alpha_1R_i + \alpha_2C_i + \alpha_3I_i + \alpha_4D_i\end{equation}
These alpha coefficients can be calculated using a training data set. The logging and analyzing system can provide the values of explanatory variables, the dependent variable $F_i$ of the training data set can be obtained by asking the user to provide based on his/her impression about document i. For example, we designate a document, then ask the user to score a grade from 1 to 10, based on how familiar s/he feels about the document. Undoubtedly, this grade is a subjective value, but it is feasible, because familiarity degree itself is a subjective value.

\subsection{How to estimate the user's familiarity degree about the target document?}
Because we are helping the user find the target document, we do not know which one it is, so we cannot use the linear regression model of section 4.1 to estimate the user's familiarity degree about the target document. The user knows which document s/he is re-finding, and knows how familiar s/he is about the target document. We can ask the user to provide his familiarity degree about the target document, but this will impose additional cognitive burden to the user and may be affected by other factors like emotions. Therefore, instead of asking the user to provide the familiarity degree about the target document, we can infer it by evaluating how the user accomplished the question and answer wizard.
Table 4 shows the variables we use to calculate the familiarity degree to the target document and their symbols.
\begin{table}
  \centering
  \begin{tabular}{|c|c|}
    \hline
    \tabhead{Variable} &
    \multicolumn{1}{|p{0.2\columnwidth}|}{\centering\tabhead{Symbol}}\\
    \hline
    Familiarity degree &\\
    to the target document & $F_t$\\
    \hline
    Average time spent per question & $T_a$\\
    \hline
    Percentage of questions &\\
    which the user has skipped & $P_s$\\
    \hline
    Percentage of questions &\\
    the user has precise answers & $P_e$  \\
    \hline
  \end{tabular}
  \caption{Variables used to estimate familiarity degree to the target.}
  \label{tab:table4}
\end{table}
We use the following linear regression formula to calculate the familiarity degree to the target document:
\begin{equation}F_t = \beta_0 + \beta_1T_a + \beta_2P_s + \beta_3P_e\end{equation}
Similarly, we can use a training data set to calculate the coefficient of each variable. However, the generation of the training data set has some difference between the previous one. We first generate a re-finding task, then ask the user to re-find it using the question and answer wizard, recording the values of the explanatory variables, if the question and answer wizard failed to locate the target file, the user can employ other methods like desktop search engine or navigation, until s/he finally located the target file, then score a familiarity degree for the target document, if s/he cannot locate the target document, just give up and generate another re-finding task.

Re-finding process is complicated process of human's interaction with computer. Every time the users access their document library, they will get some information which may affect the following re-find strategy, or may affect a user's memory about the target document.
Figure 4 shows a flow chart of generating the training data set. Letting the user first take the question wizard then use other methods can avoid affecting the user's memory about the target document.\\

\begin{figure}[!h]
\centering
\includegraphics[width=0.9\columnwidth]{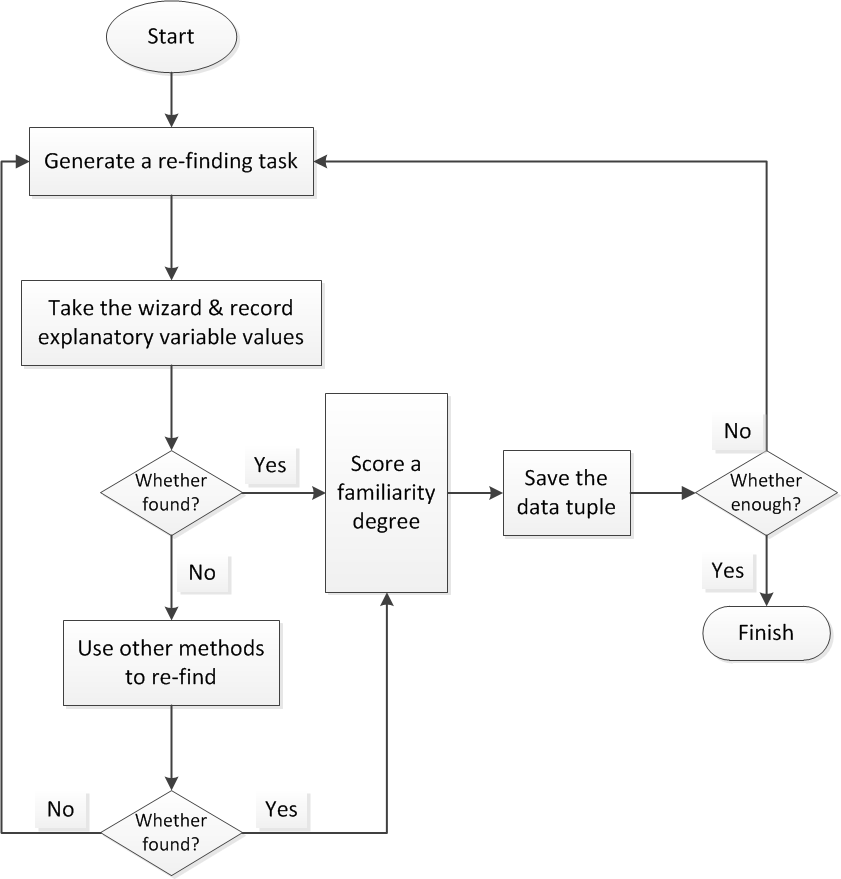}
\caption{A flow chart of generating the training data set.}
\label{fig:figure5}
\end{figure}
As we have obtained the user's familiarity degree of the target and each candidate, by computing their distances we can rank the candidates, the smaller the distance is, the higher the candidate will be ranked.

Algorithm 1 shows the pseudo-code of the NNiGN ranking algorithm using formulas (1) and (2).
\begin{algorithm}
\caption{NNiGN ranking algorithm}
\label{alg1}
\begin{algorithmic}
\STATE Let $R$ denotes the set of remaining documents after filtering
\STATE Let $d_i$ denotes document i's familiarity distance from the target document.
\STATE Use formula (2) calculate $F_t$;
\FOR{each $Doc_i\in R$}
\STATE Use formula (1) calculate $F_i$;
\STATE Let $d_i$ = $|F_i  -  F_t|$;
\ENDFOR
\STATE Sort(R); // Sort set R based on $d_i$ from small to large
\RETURN $R$;
\end{algorithmic}
\end{algorithm}

\section{User Study}

We devised a user study to evaluate how many of the re-finding tasks are difficult ones,
and how to choose parameters for recommendations, we take pages of a document as an example, tested the mean value and the median value separately.

\subsection{Generation of re-finding tasks}

To study people's re-finding behavior, the first question is how to record a re-finding process, because re-finding a file from a personal computer usually involves the user's privacy, collecting information about re-finding behaviors usually is difficult. Elsweiler and Ruthven devised a method which demands the participants to record their re-finding behaviors by keeping a dairy \cite{Elsweiler2007}, limitations of this method include lower levels of participant dedication, there is no guarantee the participants will record all the re-finding tasks they have encountered, as some re-finding tasks are so trivial that the participant may think there is no need to record it, this may bring some bias to the recorded data set.

Another way to study re-finding behavior is generating re-finding tasks artificially. Blanc-Brude et al. used a method which generates re-finding tasks by interviewing \cite{Blanc-Brude2007}. The interviewer first have a conversation with the participant, collecting information about their work and documents, then use these collected information as clues to generate re-finding tasks. This method has a limitation too, the interviewer collects information based on participants' description about their work and documents, usually the participants are unwilling to describe documents which they have a misty memory, so the generated re-finding tasks inclined to be easy tasks, which brings bias to the data set.

We propose a new method to generate re-finding tasks artificially.
Step one, the participants transfer their files which involve privacy to other places like a USB disk, then authorize the experimenter to access their personal computer, the experimenter can navigate among his/her file folders arbitrarily, then the experimenter select a set of documents randomly and take several snapshots for each of them. One of the snapshots is the first page of the document, because the user must have seen this page if s/he has ever read the document. If the document contains pictures, snapshots about a picture will also be taken, especially those pictures which are conspicuous. Since we can record a user's reading time on each page and each document\cite{liu2016method}, the pages on which the user has spent a lot of time are preferred.  Examples of the snapshots are like Figure 5 and Figure 6.
\begin{figure}[!h]
\centering
\includegraphics[width=0.9\columnwidth]{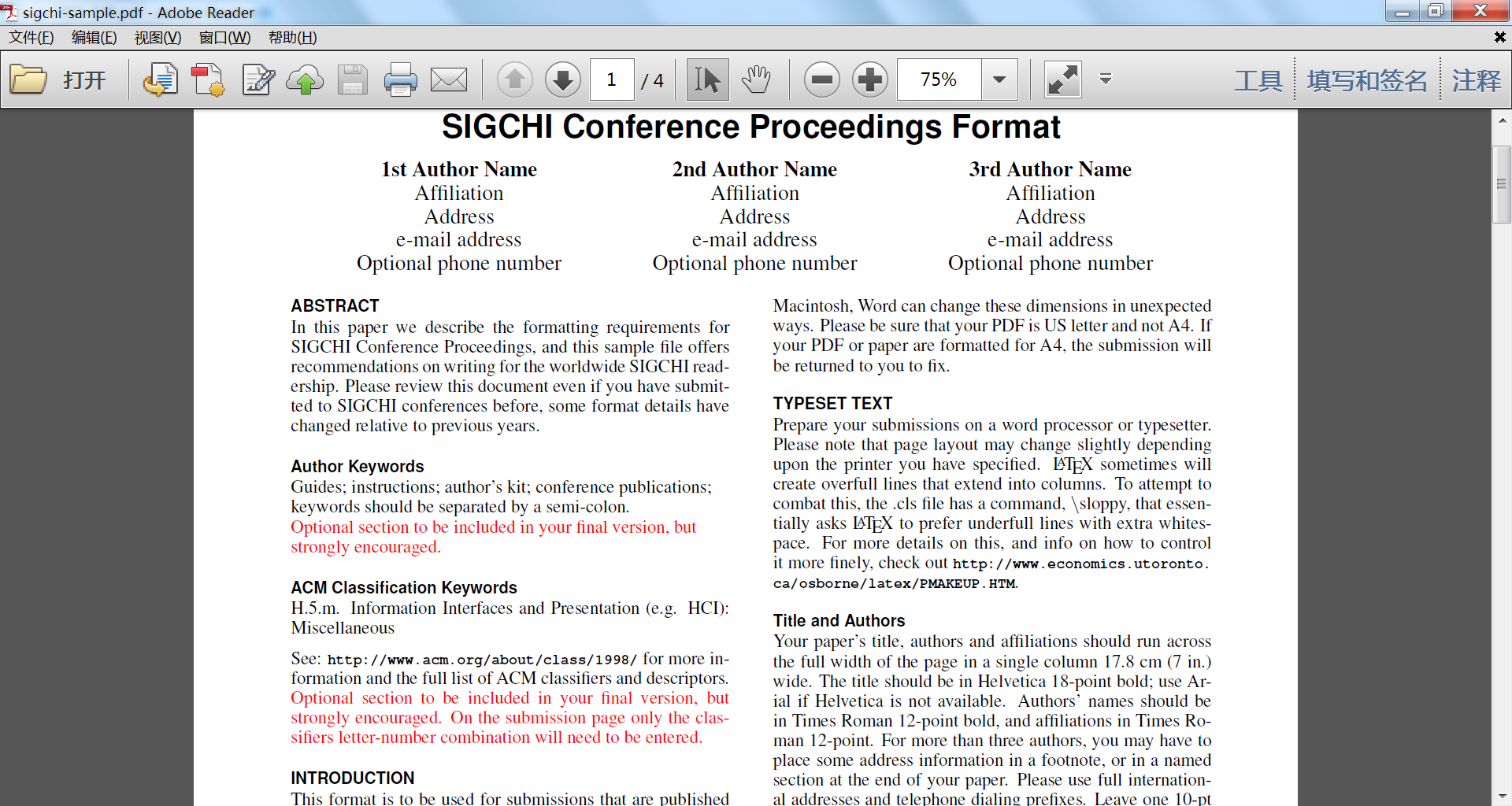}
\caption{Snapshot of the first page.}
\label{fig:figure6}
\end{figure}
\begin{figure}[!h]
\centering
\includegraphics[width=0.9\columnwidth]{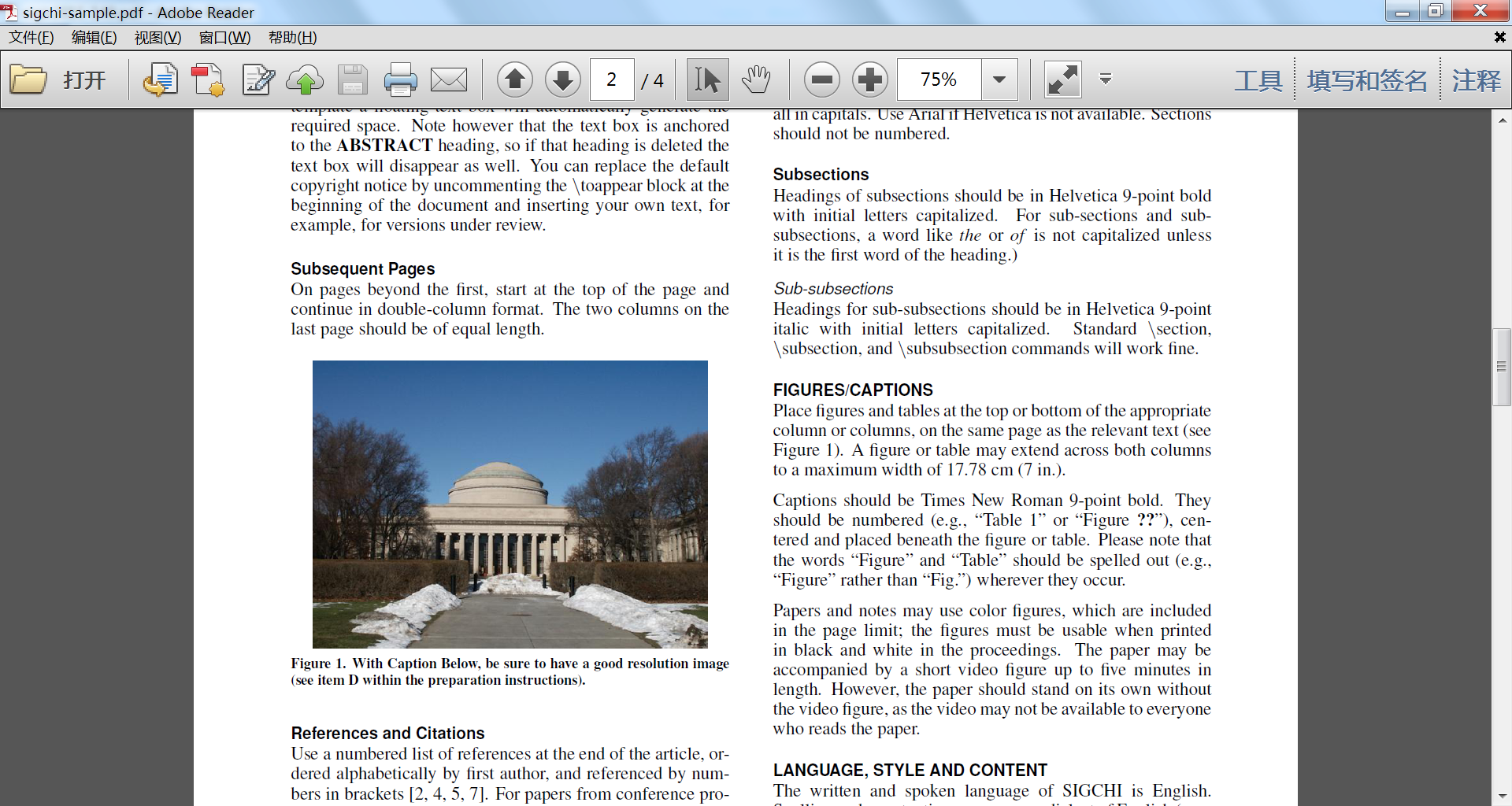}
\caption{Snapshot encircling a conspicuous picture.}
\label{fig:figure7}
\end{figure}
Step two, the experimenter shows these snapshots as a hint of the target document to the participant, then asks the participants to re-find it, meanwhile records all the needed information about the re-finding process.

How to show the snapshots to the participants is carefully considered. The hint should not give the participants any additional information like the keywords or filenames of the documents, because these information may affect the user's natural re-finding behaviors. To fulfil this demand, we take two steps. Firstly, we show the snapshots to the participant with a very short time interval, only about 0.5 sec; Secondly, we cover all titles and subtitles of the snapshots, like Figure 7, because even in a 0.5 sec of glimpse, people may catch some keywords from the titles and subtitles. Now, the participants can only use a quick glimpse of the layout of the target document to hint themselves of which document should be retrieved. This strategy brings less interferes to the use's natural re-finding behaviors, because it gives little description about the target, and will not leak any keywords of the target document.
\begin{figure}[!h]
\centering
\includegraphics[width=0.9\columnwidth]{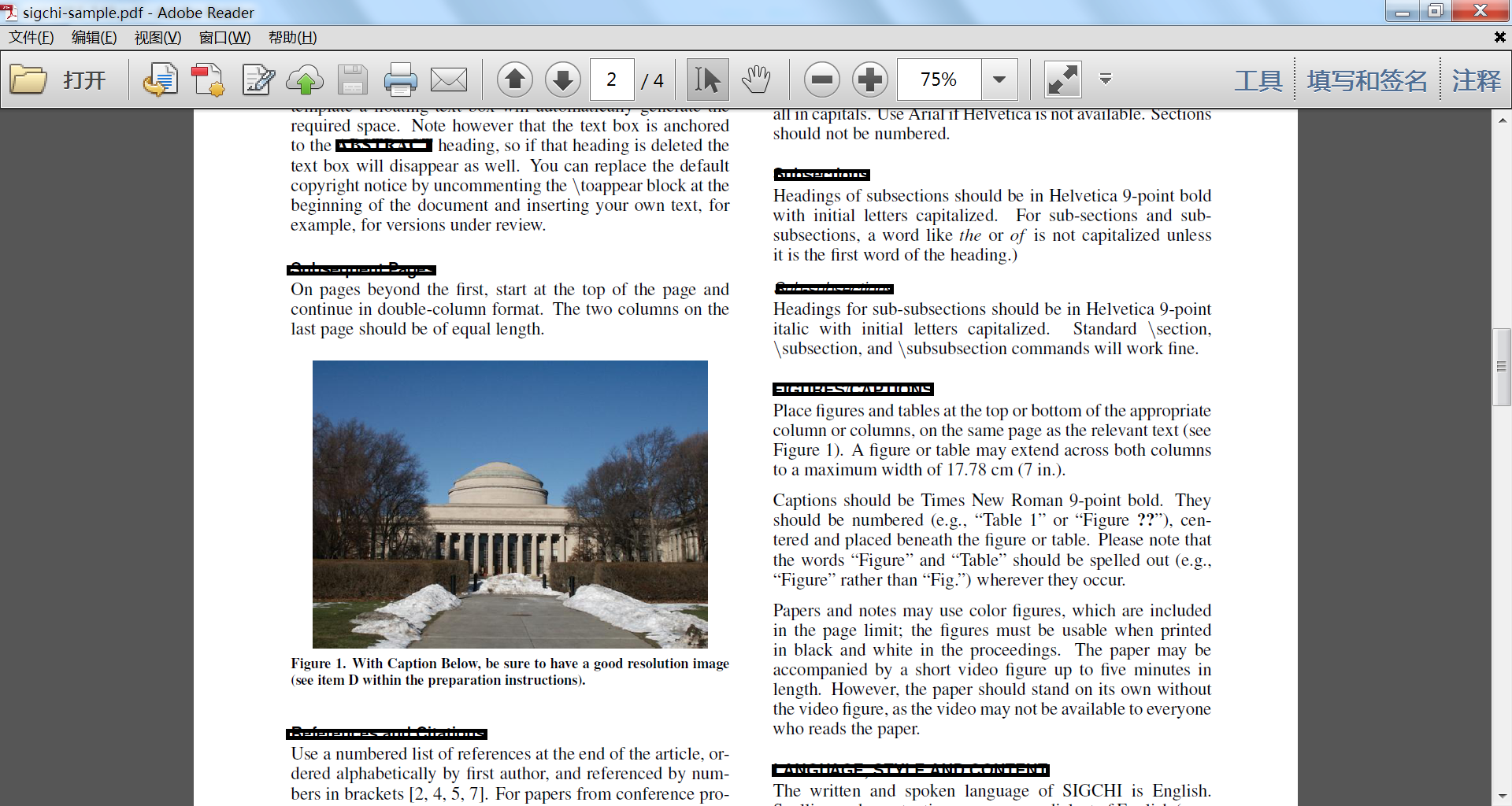}
\caption{Snapshot with prominent keywords being covered .}
\label{fig:figure8}
\end{figure}

\subsection{Definition of a difficult re-finding task}

To evaluate how many of the re-finding tasks are difficult ones, we should first define which kind of re-finding tasks are difficult re-finding tasks.

\textbf{Definition:}

    If a user cannot re-find a personal file in time $T$ by using whatever methods s/he can take, except seeking help from other people or searching from the Internet, it is a difficult re-finding task.

Because the problem we are trying to solve is re-finding from a personal file library, seeking help from others and searching from the Internet are not counted as legitimate re-finding methods. In our user study, we set $T$ equals 5 minutes.
\subsection{Results}
Because of privacy issues, it is difficult to recruit participants for out user study, finally we recruited 5 participants. All of them are research students, 3 male and 2 female, aged from 24 to 38.

\subsubsection{Percentage of difficult re-finding tasks}
We used the method mentioned previously to generate re-finding tasks, based on the definition of difficult re-finding tasks, Table 5 shows the percentage of difficult re-finding tasks among all tasks for each participant. On average, 19.6\% of re-finding tasks are difficult ones.
\begin{table}
  \centering
  \begin{tabular}{|c|c|c|}
    \hline
    \tabhead{Participants} &
    \multicolumn{1}{|p{0.3\columnwidth}|}{\centering\tabhead{Total re-finding tasks}} &
    \multicolumn{1}{|p{0.4\columnwidth}|}{\centering\tabhead{Percentage of difficult tasks}}\\
    \hline
    P1 &30  &13\%\\
    \hline
    P2 & 28 &18\%\\
    \hline
    P3 & 20 &25\%\\
    \hline
    P4 & 25 &28\%\\
    \hline
    P5 & 28 &14\%\\
    \hline
    Average &  &19.6\%\\
    \hline
  \end{tabular}
  \caption{Percentage of difficult re-finding tasks.}
  \label{tab:table5}
\end{table}

\subsubsection{Selection of classifying parameters for recommendations}
We take pages of documents as an example, record each document's pages, calculate the mean values and median values of the pages, then use these two values separately as a classifying parameter for recommendations, evaluate their performance as parameters. For example, we calculated the mean value of document pages is $53$ and the median value is $11$ for a participant, then we divide the generated re-finding tasks into two groups, in one group, we ask the participant:

\textit{``How many pages do you remember the document has?"}

with the following recommendations:

\textit{``A. More than 53 pages \quad B. Less than or equal to 53 pages"}

in the other group, we recommend the participant with the following:

\textit{``A. More than 11 pages \quad B. Less than or equal to 11 pages"}

Then we estimate the better classifying parameter by counting the participant's percentage of correct answers.

Table 6 shows the percentage of correct answers based on the two parameters for each participant.
\begin{table}
  \centering
  \begin{tabular}{|c|c|c|}
    \hline
    \tabhead{Participants} &
    \multicolumn{1}{|p{0.3\columnwidth}|}{\centering\tabhead{Use mean as parameter}} &
    \multicolumn{1}{|p{0.3\columnwidth}|}{\centering\tabhead{Use median as parameter}}\\
    \hline
    P1 & 60\% &27\%\\
    \hline
    P2 & 86\% &64\%\\
    \hline
    P3 &  70\%&80\%\\
    \hline
    P4 & 77\% &58\%\\
    \hline
    P5 & 79\% &86\%\\
    \hline
    Average &74.4\%  &63\%\\
    \hline
  \end{tabular}
  \caption{Percentage of questions answered correctly with different parameters.}
  \label{tab:table7}
\end{table}
From the data, we can conclude that the mean value of the pages serves as a better classifying parameter.
%
%
%
%
%

\bibliographystyle{acm-sigchi}
\bibliography{my}
\end{document}